\documentclass[11pt,twoside]{article}

\usepackage{asp2004}
\usepackage{epsf}
\usepackage{epsfig}
\usepackage{lscape}
\usepackage{natbib}

\begin{document}
\title{Spectroscopic Analysis of subluminous B Stars in Binaries with compact Companions}
\author{S. Geier$^1$, C. Karl$^1$, H. Edelmann$^2$, U. Heber$^1$ and R. Napiwotzki$^3$}   
\affil{$^1$ Dr.--Remeis--Sternwarte, Institute for Astronomy, University Erlangen-Nuremberg, Sternwartstr. 7, 96049 Bamberg, Germany\\
$^2$ McDonald Observatory, University of Texas at Austin,
     1 University Station, C1402, Austin, TX 78712-0259, USA\\
$^3$ Centre of Astrophysics Research, University of Hertfordshire, College        Lane, Hatfield AL10 9AB, UK}


\begin{abstract} 
The masses of compact objects like white dwarfs, neutron stars and black holes are fundamental to astrophysics, but very difficult to measure. We present the results of an analysis of subluminous B (sdB) stars in close binary systems with unseen compact companions to derive their masses and clarify their nature.  Radial velocity curves were obtained from time resolved spectroscopy. The atmospheric parameters were determined in a quantitative spectral analysis. Based on high resolution spectra we were able to measure the projected rotational velocity of the stars with high accuracy. The assumption of orbital synchronization makes it possible to constrain inclination angle and companion mass of the binaries. Five invisible companions have masses that are compatible with that of normal white dwarfs or late type main sequence stars. But four sdBs have very massive companions like heavy white dwarfs ($>\,1\,M_{\odot}$), neutron stars or even black holes. Such a high fraction of massive compact companions is not expected from current models of binary evolution.
\end{abstract}


\section{Introduction}   

The mass of a star is it's most fundamental parameter. However, a direct measurement is possible in some binary stars only. Eclipsing, double lined systems are first choice. White dwarfs, neutron stars and stellar black holes are the aftermath of stellar evolution. In binaries such faint, compact objects are outshined by their bright companions and therefore their orbital motion cannot be measured. As a consequence only lower limits to the companion mass can be derived. With the analysis method shown here, these limitations can partly be overcome.\\
Subluminous B stars, which are also known as hot sudwarf stars (sdBs), show the same spectral characteristics as main sequence stars of spectral type B, but are much less luminous. They are considered to be helium core burning stars with very thin hydrogen envelopes and masses around $0.5\,M_{\rm \odot}$.
Different formation channels have been discussed. As it turned out, a large fraction of the sdB stars are members of short period binaries \citep{maxted}. For these systems common envelope ejection is the most probable formation channel \citep{han}. In this scenario two main sequence stars of different masses evolve in a binary system. The heavier one will first reach the red giant phase and fill its Roche lobe. If the mass transfer to the companion is dynamically unstable, a common envelope is formed. Due to friction the two stellar cores loose orbital energy, which is deposited within the envelope and leads to a shortage of the binary period. Eventually the common envelope is ejected and a close binary system is formed, which contains a helium core burning sdB and a main sequence companion. When the latter reaches the red giant branch, another common envelope phase is possible and can lead to a close binary with a white dwarf companion and an sdB. All known companions of sdBs in such systems are white dwarfs or late type main sequence stars. If massive stars are involved, the primary may evolve into a neutron star (NS) or a black hole (BH) rather than a white dwarf. Since massive stars are very rare, only few sdB+NS or sdB+BH systems are expected to be found.\\
Since the spectra of the program stars are single-lined,
 they reveal no information about the orbital motion of the
 sdBs' companions, and 
 thus only their mass functions can be calculated.

 \begin{equation}
 \label{equation-mass-function}
 f_{\rm m} = \frac{M_{\rm comp}^3 \sin^3i}{(M_{\rm comp} +
   M_{\rm sdB})^2} = \frac{P K^3}{2 \pi G} .
 \end{equation}

Although the RV semi-amplitude $K$ and the period $P$ are determined
 by the RV curve, $M_{\rm sdB}$, $M_{\rm comp}$ and $\sin{i}$ remain
 free parameters.
Binary population synthesis models \citep{han} indicate a possible mass range of $M_{\rm sdB}$\,=\,0.30$-$0.48\,M$_{\rm \odot}$ for sdBs in binaries, which underwent the common envelope ejection channel. The mass distribution shows a sharp peak at about $0.46\,M_{\rm \odot}$. \\
For close binary systems, the components' stellar rotational velocities are considered to be tidally locked to their orbital motions, which means that the orbital period of the system equals the rotational period of the companions. If the companions are synchronized in this way the rotational velocity $v_{\rm rot}$ can be calculated.

\begin{equation}
v_{\rm rot} = \frac{2 \pi R_{\rm sdB}}{P} .
\end{equation}

The stellar radius $R$ is given by the mass radius relation.

\begin{equation}
R = \sqrt{\frac{M_{\rm sdB}G}{g}}
\end{equation}

The measurement of the projected rotational velocities 
 $v_{\rm rot}\,\sin\,i$ 
therefore allows to constrain the systems' inclination angles $i$.
With $M_{\rm sdB}$ as free parameter the mass function can be solved and the inclination angle as well as the companion mass can be derived. Because of $\sin{i} \leq 1$ a lower limit for the sdB mass is given.
To constrain the system parameters in this way it is necessary to measure $K$, $P$, $\log{g}$ and $v_{\rm rot}\sin{i}$ with high accuracy. 

\section{Observations and Radial Velocity Curves}

Ten stars were observed at least twice 
 with the high resolution spectrograph UVES at the ESO\,VLT.
Additional observations were made 
 at the ESO\,NTT (equipped with EMMI), the Calar Alto Observatory 3.5\,m telescope
 (TWIN) and the 4\,m WHT (ISIS) at La Palma.
Two of the stars (PG\,1232$-$136, TONS\,183) were observed with the high resolution FEROS instrument at the 2.2\,m ESO telescope at La Silla.\\
Radial velocities of the individual observations were determined
 by calculating the
 shifts of the measured wavelengths relative to their laboratory values. Results are displayed in Table~1.

\section{Atmospheric parameters and projected rotational velocities}

The spectra were corrected for the measured RV and coadded. Atmospheric parameters were
 determined by fitting simultaneously each observed hydrogen and helium line
 with a grid of metal-line blanketed LTE model spectra. Results are listed in Table~1.

\begin{table} 
\label{param}
\caption{Effective temperatures, surface gravities, orbital periods, radial velocity semi-amplitudes and projected rotational velocities
of the visible components.
Typical error margins for  $T_{\rm eff}$ and $\log g$ are
500\,K and 0.05\,dex respectively.
$\dag$ Orbital parameters of this system are still preliminary.}
\smallskip
\begin{center}

\begin{tabular}{lcccccc}

\tableline
\noalign{\smallskip}
System & $T_{\rm eff}$ & $\log{g}$ & $P$ & $K$ &  $v_{\rm rot}\,\sin\,i$ \\
       & [K] &  & [d] & [${\rm km\,s^{-1}}$] & [${\rm km\,s^{-1}}$] \\ 
\noalign{\smallskip}
\tableline
\noalign{\smallskip}
HE\,0532$-$4503 & 25390 & 5.32 & 0.26560 $\pm$ 0.00010 & 101.5 $\pm$ 0.2 & 11.1 $\pm$ 0.6 \\
PG\,1232$-$136 & 27500 & 5.62 & 0.36300 $\pm$ 0.00030 & 129.6 $\pm$ 0.04 & 6.2 $\pm$ 0.8 \\
WD\,0107$-$342$\dag$ & 24300 & 5.32 & 0.38000 $\pm$ 0.00050 & 135.0 $\pm$ 1.0 & 20.4 $\pm$ 0.9 \\
HE\,0929$-$0424 & 29470 & 5.71 & 0.44000 $\pm$ 0.00020 & 114.3 $\pm$ 1.4 & 7.1 $\pm$ 1.0 \\
HE\,0230$-$4323 & 31100 & 5.60 & 0.44300 $\pm$ 0.00050 & 64.1 $\pm$ 1.5 & 12.7 $\pm$ 0.7 \\
TONS\,183 & 26100 & 5.20 & 0.82770 $\pm$ 0.00020 & 84.8 $\pm$ 1.0 & 6.7 $\pm$ 0.7 \\
HE\,2135$-$3749 & 30000 & 5.84 & 0.92400 $\pm$ 0.00030 & 90.5 $\pm$ 0.6 & 6.9 $\pm$ 0.5 \\
HE\,1421$-$1206 & 29570 & 5.55 & 1.18800 $\pm$ 0.00100 & 55.5 $\pm$ 2.0 & 6.7 $\pm$ 1.1 \\
HE\,1047$-$0436 & 30242 & 5.66 & 1.21325 $\pm$ 0.00001 & 94.0 $\pm$ 3.0 & 6.2 $\pm$ 0.6 \\
HE\,2150$-$0238 & 30200 & 5.83 & 1.32090 $\pm$ 0.00500 & 96.3 $\pm$ 1.4 & 8.3 $\pm$ 1.3 \\
HE\,1448$-$0510 & 34690 & 5.59 & 7.15880 $\pm$ 0.01300 & 53.7 $\pm$ 1.1 & 6.7 $\pm$ 2.5 \\
WD\,0048$-$202  & 29960 & 5.50 & 7.44360 $\pm$ 0.01500 & 47.9 $\pm$ 0.4 & 7.2 $\pm$ 1.3 \\
\hline
\\
\end{tabular}

\end{center}
\end{table}

In order to derive $v_{\rm rot}\,\sin\,i$,
 we compared the observed spectra 
 with rotationally broadened, synthetic line profiles.
The latter ones were computed using the LINFOR program. Since sharp metal lines
 are much more sensitive to rotational broadening
 than Balmer or helium lines, all visible metal lines were included. A simultaneous fit of elemental abundance and projected rotational velocity was performed separately for every identified line using the FITSB2 routine \citep{napiwotzki}. Mean value and standard deviation were calculated from all measurements.
Seeing induced variations in the instrumental profile and the noise level were the dominant error sources. Information on the actual seeing conditions for every exposure have been extracted from the ESO seeing monitor archive.
All other possible sources of systematic errors turned out to be negligible.\\

\begin{figure}[t!]
 
 \plottwo{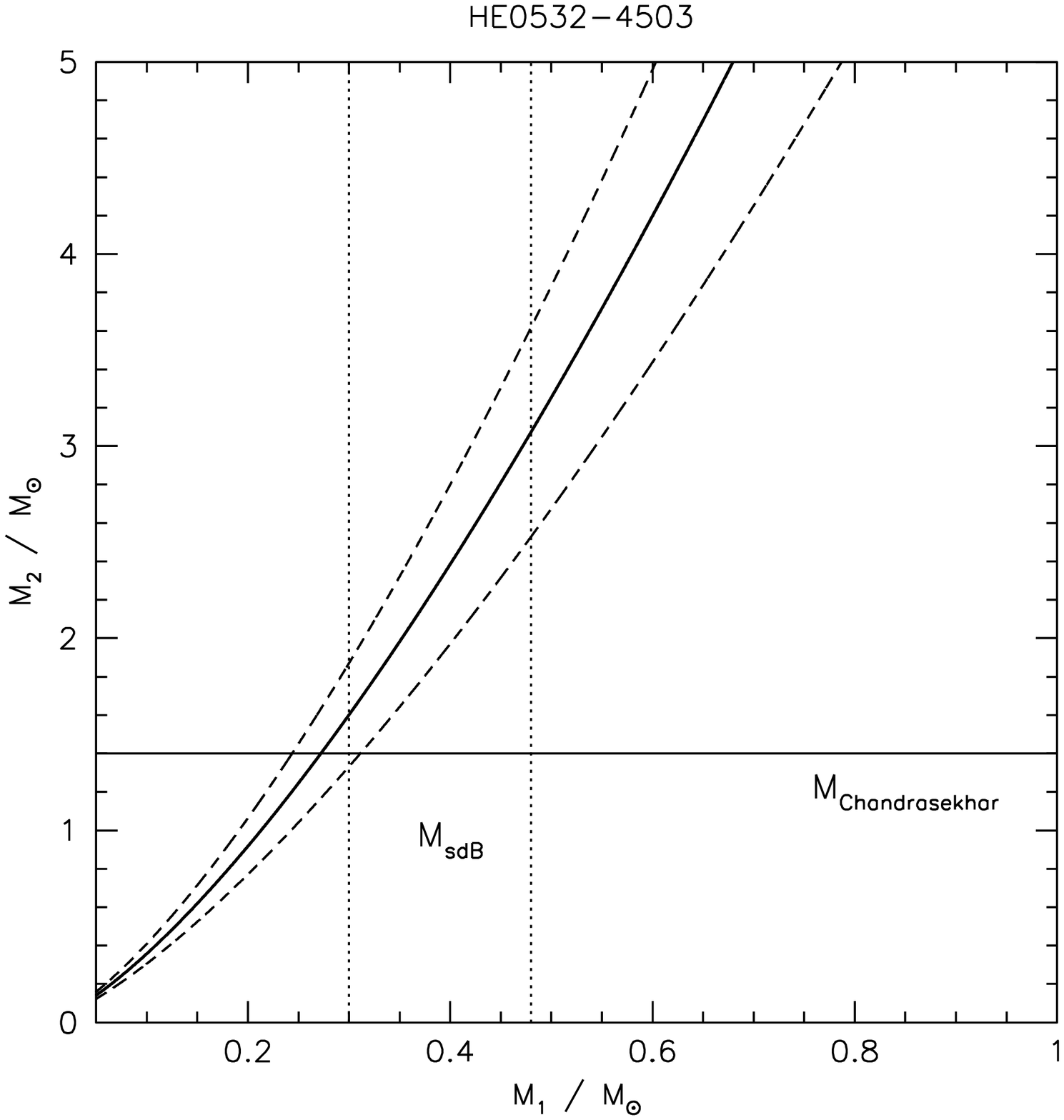}{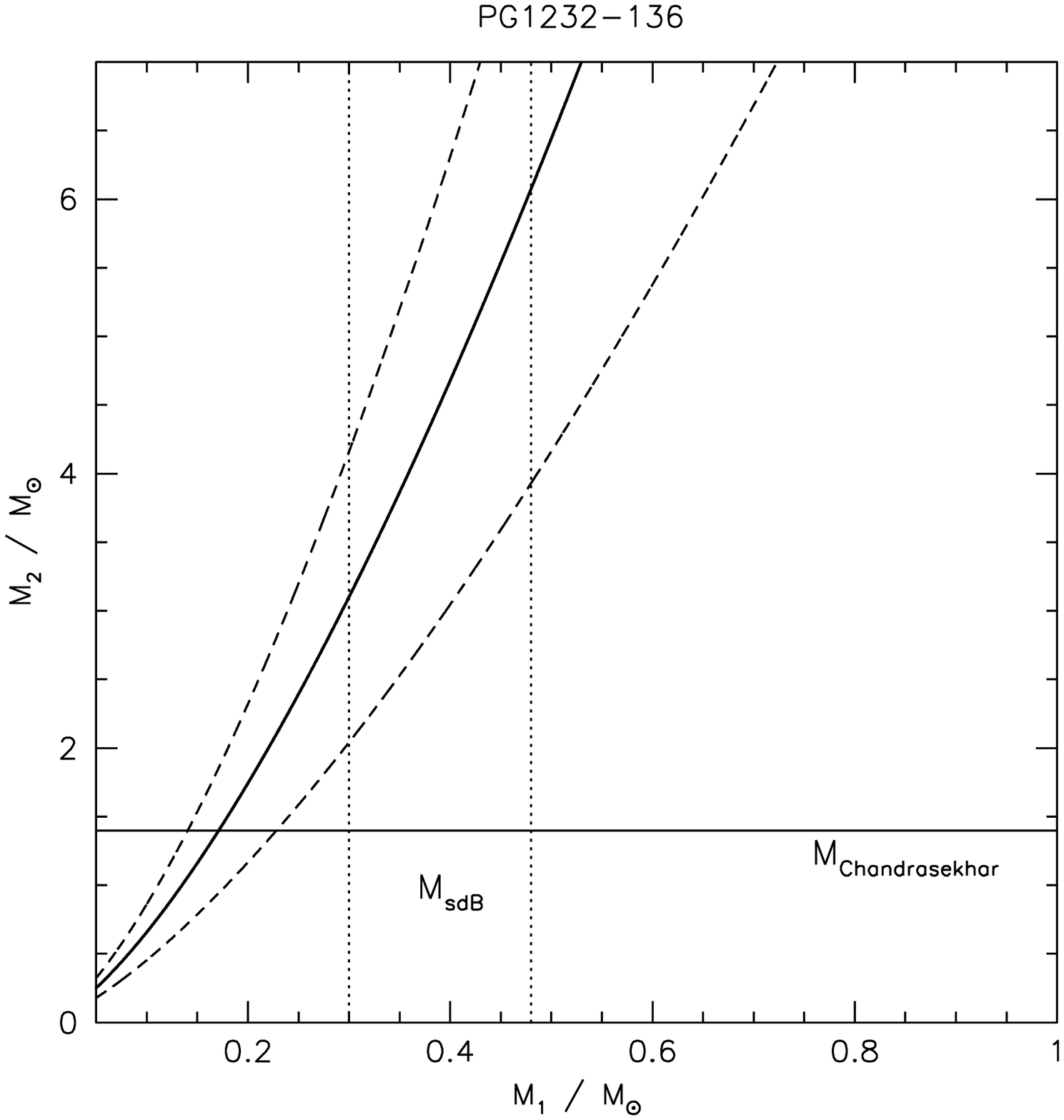}
 \caption{Companion mass as a function of primary (sdB) mass of the binaries HE\,0532$-$4503 (left hand panel) and PG\,1232$-$136 (right hand panel). The horizontal line marks the Chandrasekhar limit. The dotted vertical lines mark the theoretical sdB mass range for the common envelope ejection channel \citep{han}.}

 \label{lock1}
\end{figure}




\section{Nature of the unseen Companions}

Knowing $P$, $K$, $log\,g$ and $v_{rot}\sin{i}$ we can calculate the mass of the unseen companion from equations (1) - (3) for any given primary mass (see Fig. \ref{lock1}). We adopt the mass range from \citep{han}, marked by the dotted lines in Fig. \ref{lock1}.
To constrain the companion masses and inclination angles we adopted the mass range for sdBs from \citet*{han}. There are no spectral signatures of companions visible. Main sequence stars with masses higher than $0.45\,M_{\rm \odot}$ could therefore be excluded because of their high luminosities in comparison to the sdB stars. The possible companion masses can be seen in Tab.~2. Four of the analysed systems have companion masses, which are compatible with either typical white dwarfs (WD) or late main sequence stars (late MS). The companion of WD\,0107$-$0342 is a white dwarf of intermediate mass.\\
The very similar HE\,0929$-$0424 and TONS\,183 have to have quite massive companions. If their sdB primaries are of low mass ($\approx \,0.3\,M_{\odot}$) the companions could be a heavy white dwarfs. At the most probable sdB mass, however, their companions would exceed the Chandrasekhar mass limit. There are only two kinds of objects known with such high masses and such low luminosities - neutron stars (NS) and stellar mass black holes (BH). The two systems HE\,0532$-$4503 and PG\,1232$-$136 have even higher companion masses, which would exceed the Chandrasekhar limit in any case (see Fig.~\ref{lock1}). The three binaries HE\,1448$-$0510, HE\,2150$-$0238 and WD\,0048$-$202 could not be solved with the described method. The minimum sdB masses exceeded $1.3\,M_{\rm \odot}$ and were not consistent with the theoretical mass range. These systems cannot be synchronized.\\
Binaries hosting a neutron star or a black hole are a very rare class of objects. From about 50 analysed sdB binaries in our samples, we found two, possibly four candidate systems. This fraction of 4-8 \% is much too high to be compatible with any binary evolution model known so far (Podsiadlowski priv. comm.).\\
The reliability of our method has to be discussed again. Only the high precision of the $v_{\rm rot}\sin{i}$ measurements made it possible to constrain the parameters reasonably. The projected rotational velocities were low and very close to the detection limit. As described above we tried to quantify all possible systematic effects and the overall results were very consistent. Only slight changes in $v_{\rm rot}\sin{i}$ would lead to inconsistent solutions, because the method is very sensitive to this parameter. But even if there would be unaccounted systematic effects (e.g. short period pulsations), they would always cause an extra broadening of the lines. The measured broadening is then due to rotation and the unaccounted effects, which means the deduced rotational broadening is higher than the real one. All systematic effects lead to lower $v_{\rm rot}\sin{i}$. But the unexpectedly high masses of the companions are caused by the unexpectedly low measured projected rotational velocities as can be seen from the equations (1) - (3). Unaccounted systematic effects would therefore lead to even higher companion masses.\\
Our method rests on the assumption of orbital synchronization. 
Many eclipsing binaries in close orbits were observed to be synchronized. At least two short period ($\approx \,0.1\,d$) sdB binaries with white dwarf companions are proven to be synchronized \citep[][Geier et al. these proceedings]{orosz}.\\
The question, which mechanism is responsible for this effect, is not yet settled.  The approximation formulas for the synchronization time of \citet*{zahn} and \citet*{tassoul} for stars with radiative envelopes and convective cores were used to compare with the lifetime on the Extreme Horizontal Branch (EHB) $t_{\rm EHB} \approx 10^{8}\,{\rm yr}$. Nine short period binaries can be solved consistently and have synchronization times much lower than their EHB lifetimes. Three systems could not be solved consistently. These three binaries have the longest orbital periods and synchronization times near or above the EHB lifetime. The two systems with the most massive companions have the shortest orbital periods and therefore should most likely be synchronized (see Table 1 \& 2).\\
But as long as the question of tidal synchronization is not settled  \citep[see review by][]{zahn2}, all timescales have to be taken with caution. Detailed calculations, which take the internal structure of sdBs into account, are not available and urgently needed.
From our results we draw the conclusion, that the assumption of tidally locked rotation is reasonable in the case of close sdB binaries for orbital periods shorter than $1.3\,d$.

\section{Conclusion}

Out of 12 analysed sdB binaries, five have companion masses compatible with white dwarfs of typical masses or late type main sequence stars. These systems are in full agreement with binary population synthesis simulations. Four binaries have surprisingly high companion masses, which leads to the conclusion, that the companions have to be white dwarfs of unusually high mass or even neutron stars or black holes. This high fraction cannot be explained with current evolutionary calculation. As our analysis assumes synchronization a better understanding of this process for sdB stars is urgently needed. More observations of close sdB binaries and candidate systems with high and low resolution spectrographs are needed to enhance the size of our sample.\\
The presence of such a high fraction of heavy binaries in our samples raises several questions. Is the formation and evolution of sdB stars linked to that of heavy compact objects like neutron stars or black holes? Can sdB stars be used as tracers to find more of these exotic objects? Is there a hidden population of these objects, which is not visible through strong X-ray emission?\\

\begin{table} 
\label{comp}
\caption{Inclination angles, rotational velocities, companion masses and possible nature of the unseen companions. The lower companion mass corresponds to an sdB of $0.3\,M_{\odot}$, the higher limit to an sdB of $0.48\,M_{\odot}$
$\dag$ Derived parameters of this system are still preliminary.}
\smallskip
\begin{center}
\begin{tabular}{lcccccc}

\tableline
\noalign{\smallskip}

System &  $i$ & $v_{\rm rot}$ & $M_{\rm comp}$ & Companion\\
       & [deg] & [${\rm km\,s^{-1}}$] & [$M_{\rm \odot}$] \\ 
\noalign{\smallskip}
\tableline
\noalign{\smallskip}
HE\,0532$-$4503 & 13 $-$ 17 & 47 & 1.40 $-$ 3.60 & \bf NS/BH\\
PG\,1232$-$136 & 14 $-$ 19 & 25 & 2.00 $-$ 7.00 & \bf NS/BH \\
WD\,0107$-$342$\dag$ & 17 $-$ 22 & 33 & 0.45 $-$ 0.95 & WD \\
HE\,0929$-$0424 & 23 $-$ 29 & 18 & 0.60 $-$ 2.40 & \bf WD/NS/BH \\
HE\,0230$-$4323 & 38 $-$ 50 & 21 & 0.18 $-$ 0.35 & WD/late MS\\
TONS\,183 & 22 $-$ 29 & 18 & 0.60 $-$ 2.40 & \bf WD/NS/BH \\
HE\,2135$-$3749 & 66 $-$ 90 & 8 & 0.35 $-$ 0.45 & WD/late MS\\
HE\,1421$-$1206 & 56 $-$ 90 & 8 & 0.15 $-$ 0.30 & WD/late MS\\
HE\,1047$-$0436 & 62 $-$ 90 & 7 & 0.35 $-$ 0.60 & WD/late MS\\
HE\,2150$-$0238 & -- & -- & -- & no solution \\
HE\,1448$-$0510 & -- & -- & -- & no solution \\
WD\,0048$-$202  & -- & -- & -- & no solution \\
\hline\\
\end{tabular}

\end{center}

\end{table}

\end{document}